\documentclass[structabstract]{aa}  

\usepackage{graphicx}

\usepackage{txfonts}

\usepackage{natbib} 

\bibpunct{(}{)}{;}{a}{}{,} 

\begin{document}

   \title{Neutral gas in Lyman-alpha emitting galaxies Haro 11 and ESO 338-IG04 measured through sodium absorption\thanks{Based on observations made with ESO Telescopes at the Paranal Observatory under programme IDs 083.B-0470 and 60.A-9433}}

  \authorrunning{Sandberg et al. 2013}
  \titlerunning{Na D in Haro 11 and ESO 338-IG04}

   \author{A. Sandberg\inst{1}, G. \"Ostlin\inst{1},  M. Hayes\inst{2,3}, K. Fathi\inst{1}, D. Schaerer\inst{4,3}, J.M. Mas-Hesse\inst{5}, T. Rivera-Thorsen\inst{1}            
          }

   \institute{The Oskar Klein Centre, Department of Astronomy, Stockholm University, AlbaNova, 106 91 Stockholm, Sweden.\\
   \email{sandberg@astro.su.se}
           \and
              Universit\'e de Toulouse; UPS-OMP; IRAP; Toulouse, France
              \and
              CNRS; IRAP; 14, avenue Edouard Belin, F-31400 Toulouse, France
              \and
              Observatoire de Gen{\`e}ve, Universit{\'e} de Gen{\`e}ve, 51 Ch. des Maillettes, 1290 Versoix, Switzerland            
              \and
              Centro de Astrobiolog\'{\i}a (CSIC--INTA), Madrid, Spain
             }

 
  \abstract
   {The Lyman alpha emission line of galaxies is an important tool for finding galaxies at high redshift, and thus probe the structure of the early universe. However, the resonance nature of the line and its sensitivity to dust and neutral gas is still not fully understood.}
   {We present measurements of the velocity, covering fraction and optical depth of neutral gas in front of two well known local blue compact galaxies that show Lyman alpha in emission: ESO 338-IG 04 and Haro 11. We thus test observationally the hypothesis that Lyman alpha can escape through neutral gas by being Doppler shifted out of resonance. }
   {We present integral field spectroscopy from the GIRAFFE/Argus spectrograph at VLT/FLAMES in Paranal, Chile. The excellent wavelength resolution allows us to accurately measure the velocity of the ionized and neutral gas through the H$\alpha$ emission and Na D absorption, which traces the ionized medium and cold interstellar gas, respectively. We also present independent measurements with the VLT/X-shooter spectrograph which confirm our results.}
   {For ESO 338-IG04, we measure no significant shift of neutral gas. The best fit velocity is -15$\pm$16 km/s. For Haro 11, we see an outflow from knot B at 44$\pm$13 km/s and infalling gas towards knot C with 32$\pm$12 km/s. Based on the relative strength of the Na D absorption lines, we estimate low covering fractions of neutral gas (down to 10\%) in all three cases.}
   {The Na D absorption likely occurs in dense clumps with higher column densities than where the bulk of the Ly~$\alpha$ scattering takes place. Still, we find no strong correlation between outflowing neutral gas and a high Lyman alpha escape fraction.  The Lyman alpha photons from these two galaxies are therefore likely escaping due to a low column density and/or covering fraction.}


   \maketitle

\section{Introduction}

The Lyman alpha (Ly $\alpha$) emission line was suggested as a probe for discovering high redshift galaxies already by \citet{partridge-peebles1967}. When absorbed by neutral hydrogen gas, approximately two thirds of the ionizing photons from hot, massive stars are reprocessed into Ly $\alpha$ photons following case B recombination. The fraction of the bolometric flux contained in the Ly $\alpha$ line should be as high as 6-7 per cent in a young, star-forming stellar population.
However, early surveys designed for finding Lyman alpha emitting galaxies (LAEs) came up blank. \citep[see e.g.][and references therein]{pritchet1994}. 

It was only with deeper and larger surveys \citep[e.g.][]{cowie-hu1998} that targeting Ly $\alpha$ became the successful method that it is today for finding galaxies at redshifts $z \gtrsim 2$. It has since then been widely used for mapping out the large-scale structure of the high redshift universe \citep[e.g.][]{rhoads-2000, kudritzki-2000, malhotra-rhoads2002, ouchi-2003, ouchi-2005, gawiser-2006, ajiki-2006, gronwall-2007, pirzkal-2007, finkelstein-2008, nilsson-2009, yuma-2010, ouchi-2010,guaita-2011,adams-2011,blanc-2011,shibuya-2012} and even constraining the epoch of cosmic reionization \citep[e.g.][]{malhotra-rhoads2004,dijkstra-2007,ono-2010,jensen-2012}. 

As shown by \citet{hayes-2010}, up to 90 percent of star-forming galaxies in high redshift surveys emit too little Ly $\alpha$ to be detected by standard criteria. The first candidate to be blamed for this apparent discrepancy between theory and observation was absorption by dust, which is prominent in the ultraviolet. Early surveys in the local Universe hinted at an anti-correlation between the metallicity (which generally correlates with dust content) and Ly $\alpha$ luminosity \citep[e.g.][]{meier-terlevich1981},  but it soon became clear that dust alone could not explain the deviation from recombination theory \citep{giavalisco-1996}. In particular one galaxy, I Zw 18, showed a very low metallicity combined with strong Ly $\alpha$ absorption \citep{kunth-1994}.

The attention then turned to the resonant scattering of Ly $\alpha$ in H {\sc i}, which had been explored theoretically for a time \citep[e.g.][]{osterbrock1962,adams1972}. If the path length of the Ly $\alpha$ photons is greatly increased in multiple scatterings on the way through the galaxy, even small amounts of dust can cause a large absorption \citep{neufeld1990}.
However, if the neutral gas is shifted in velocity, the Ly $\alpha$ photons are shifted out of resonance and can escape from the galaxy more easily.
\citet{kunth-1998} showed from a sample of eight local gas-rich dwarf galaxies that the emission of Ly $\alpha$ in each case exhibited a P Cygni profile accompanied by blueshifted low ionization state (LIS) absorption lines, suggesting that an outflow of neutral gas would allow the Ly $\alpha$ emission to escape towards us. Modern simulations \citep[e.g.][Duval et al. 2012, submitted]{verhamme-2006, garel-2012} thus take into account a combination of the relative velocities of ionized and neutral hydrogen gas, as well as dust, and their respective morphologies, that all govern the escape of Ly $\alpha$ emission from galaxies. 

Although neutral gas is best traced through LIS absorption lines in the UV, such a study at low redshift requires a space telescope due to ultraviolet absorption in the atmosphere. The choice is then either HST/STIS which has a low sensitivity or HST/COS with limited spatial information. However, the emission and absorption of Ly $\alpha$ may vary over small scales \citep{mas-hesse-2003, ostlin-2009} and a detailed study requires some degree of spatial resolution. It is therefore useful to identify an alternative set of longer wavelength absorption features that still form in the neutral ISM, in order to facilitate the use of contiguous integral field spectrographs attached to large-aperture telescopes on the ground. Neutral gas motions in star-forming galaxies have been studied in the past using, e.g. Mg {\sc ii} 2796,2803 \citep[e.g.][]{churchill-vogt2001,mshar-2007,martin-bouche2009,nestor-2011} and Na {\sc i} 5889,5895 \citep[e.g.][]{heckman-2000,rupke-2002,martin2005,chen-2010}, although never before have such kinematic tracers been used with reference to Ly $\alpha$. 

In this paper, we present the first spatially resolved study of absorption from Na {\sc i} in the cold interstellar gas in two nearby Ly $\alpha$ emitting blue compact galaxies. Specifically, we target the sodium resonance absorption doublet ($\lambda \lambda$ 5889.95,5895.92), which we will refer to as Na D in this paper. The galaxies are Haro 11 (ESO 350-IG38) and ESO 338-IG04 (Tololo 1924-416) which are both objects of intense study due to their Ly $\alpha$ emission and the possible similarities to LAEs and Lyman Break Galaxies (LBGs) at high redshift (see Section~\ref{Sect:Sample}). 

Na D is generally stronger than any other resonance line in the optical (such as K {\sc i} or Ca {\sc ii}, which are strongly depleted in diffuse, low-velocity clouds, see e.g. \citet{spitzer1968}). With an ionizing potential of 5.14 eV it is a good tracer of neutral hydrogen gas. By simultaneously studying the H$\alpha$ emission line, we can accurately determine the location, extent and velocity distribution of the ionized hydrogen in these galaxies. We are thus able to measure the relative velocities between the ionized and neutral hydrogen gas, which is precisely the Doppler shift relevant for Ly~$\alpha$ transmission. With integral field unit spectra from Argus at VLT/FLAMES we achieve a high spectral resolution combined with spatial information. We include an independent measurement from VLT/X-Shooter to confirm these results, and to estimate the stellar contamination of the spectra.

The paper proceeds as follows; in Section 2 we describe previous studies of the two galaxies in our sample in more detail. In Section 3 we describe our data and the reduction steps. In Section 4 we present our results. In section 5 we discuss possible complications from stellar contamination. In Section 6 we discuss our results and in Section 7 we leave our concluding remarks. 

   \begin{figure*}
   \centering
   \includegraphics[width=16cm]{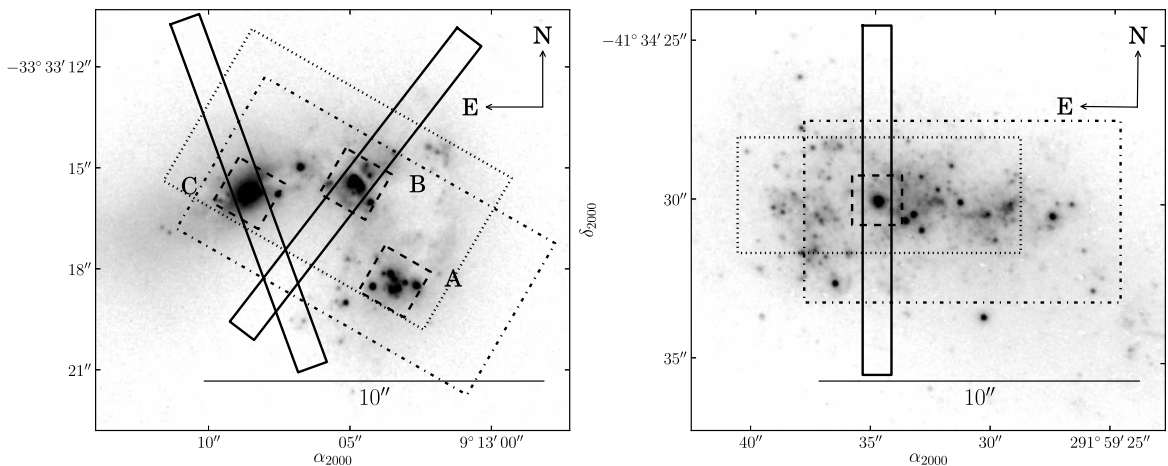}
   \caption{Slit positions, field of view (FOV) and approximate aperture positions for Haro 11 (left) and ESO 338-IG04 (right), overlayed on HST/ACS F550M continuum images (i.e. close in wavelength to the Na D feature). The X-Shooter slits are 0.9 (or 1.0 for the UVB-arm, not marked in the figure) arcsec wide and 11 arcsec long and marked with solid lines. The FOV for the LR6 wavelength range (which includes H$\alpha$) is marked with dash-dotted lines while the shorter LR5 range (including Na D) is marked with dotted lines. The Argus binned "apertures" are 1.56 arcsec wide squares marked with dashed lines. The VLT/FLAMES FOV shown here shows the area for which we have full exposure after taking into account the dithering pattern. }
              \label{Fig_Slits}
    \end{figure*}

\section{Sample}\label{Sect:Sample}

The two galaxies in our study have already been studied extensively for their interesting properties, but the neutral gas velocity as estimated from the sodium doublet has not been measured with high accuracy and spatial resolution until now.
Analyzing VLT/UVES spectra of ESO 338-IG04, \citet{ostlin-2007} estimated an outflow velocity $\sim$ 20 km/s from Na D, but emphasized that with their low signal to noise they could not estimate the relative ISM versus stellar contribution.

An HST UV imaging program including the two galaxies \citep{kunth-2003, hayes-2005, hayes-2007, ostlin-2009} gives us information on the Ly $\alpha$ and UV continuum morphologies.  

Haro 11 is a blue compact galaxy at $z=0.02$ with three large condensations or "knots".
The knots are traditionally called A, B and C \citep{vader-1993} as shown in the left panel of Figure \ref{Fig_Slits}.
Knot C appears to be the brightest knot in the ultraviolet and it also shows Ly $\alpha$ in emission, while knot B is brightest in H$\alpha$, where Ly $\alpha$ is instead absorbed. These two knots are therefore of particular interest as we try to explain why Ly $\alpha$ is seen from one region but is absorbed in the other. 
Some evidence suggests that Haro 11 might be the result of a merger of dwarf galaxies \citep{ostlin-2001}, particularly due to the irregular appearance, the high relative velocities of several hundred km/s, the broad emission lines, and the presence of a tidal arm structure 
with a high redshifted velocity relative to the rest of the system ({\"O}stlin et al. 2013, in preparation). The H $\alpha$ line width within a star forming knot is as high as $\sim$ 270 km/s (FWHM), and shows strong multi-component features. 

\citet{kunth-1998} present HST/GHRS spectra of Haro 11 around Ly $\alpha$ and the interstellar O {\sc i} ($\lambda 1302 \AA$) and Si {\sc ii} ($\lambda 1304 \AA$) lines, estimating an outflow velocity of the neutral medium of circa 60 km/s. Both absorption lines are very broad, indicating multiple components along the line of sight spanning roughly 200 km/s in velocity. The Ly $\alpha$ line profile has a strong underlying absorption, extending more than 1500 km/s on the blue side of the emission. The line does not show a clear P Cygni profile, and the underlying absorption seems to extend also to the red side. Evidence for several blueshifted Ly $\alpha$ components was found, indicating multiple absorbing gas clouds. Unfortunately, it is not clear exactly where in the galaxy the $1\farcs7 \times 1\farcs7$ GHRS aperture was pointed \citep[see discussion in][]{hayes-2007}. 

ESO 338-IG04 (or just ESO 338 as we might refer to it in this paper) has slightly narrower emission lines than Haro 11 (FWHM $\sim$ 180 km/s) and the Ly $\alpha$ emission is concentrated to a bright central region. ESO 338 is riddled with "super star clusters"; small knots of intense star formation \citep{ostlin-1998,ostlin-2003,ostlin-2007}. The ongoing starburst is about 40 Myr old and was likely triggered by a merger with a small galaxy or from an interaction with remaining debris from a previous encounter with the companion galaxy ESO 338-IG04b \citep{ostlin-2001, cannon-2004}. When we discuss ESO 338 below, we mainly focus on the bright central region, which we refer to as knot A \citep[][cf. Figure~\ref{Fig_Slits}]{hayes-2005} (also known as cluster \# 23 in the 'inner' sample of \citet{ostlin-1998}).

\citet{hayes-2005} present a detailed description of a HST/STIS long slit spectrum  across ESO 338. The spectrum shows diffuse Ly $\alpha$ emission in several regions along the slit, but over knot A the Ly $\alpha$ emission is weak and shows only a hint of blueshifted absorption. However, the low resolution of the spectrum makes a detailed kinematical study difficult, and the emission line lies very close to the strong geocoronal Ly $\alpha$ line. The slit is also only $0\farcs2$ wide, and was not optimized to be centered specifically over knot A. It is therefore not clear how well this spectrum samples the Ly $\alpha$ profile in this region. An outflow velocity from this STIS spectrum was estimated by \citet{schwartz-2006} from low ionization state absorption lines in the UV to be $47 \pm 70$ km/s.

In this paper we look mainly at the strong knots in the respective galaxies; see Figure \ref{Fig_Slits}. Even though our field of view with VLT/FLAMES is larger, the Na D signature is very weak and the areas marked are the only regions where we can successfully fit the Na D lines, even after spatially binning our spectra. 

\section{Observations and Data reduction}

\subsection{Integral Field Spectroscopy with VLT/FLAMES}

The integral field spectroscopy observations (ESO ID 083.B-0470(A)) were performed between June 9 and July 6 2009 with the Argus integral field unit of the GIRAFFE spectrograph at the FLAMES instrument at VLT/UT2. 
Argus consists of 14x22 lenslets arranged in a rectangular grid, corresponding to a field of view of approximately 7 x 11 arcseconds. Each lenslet is then connected to a fiber which leads the light to a spectrograph. One spectrum for each lenslet is thus obtained.

We observed the two galaxies Haro 11 and ESO 338-IG04, both with two "low resolution" ({$\mathcal{R} = 11800 - 13700$}) gratings. The LR6 grating (which covers $\lambda$ 6438-7184 $\AA$) was used to observe the H$\alpha$ emission line ($\lambda$ 6562.82 $\AA$), and the LR5 grating ($\lambda$ 5741-6524 $\AA$) at the Na D doublet ($\lambda \lambda$ 5889.95,5895.92 $\AA$).

The exposure time was $195 \textrm{s} \times 8 = 1560 \textrm{s}$ for the LR6 grating and $895 \textrm{s} \times 12 + 915 \textrm{s} \times 2 = 12570 \textrm{s}$ for LR5 for both galaxies. The seeing was $0\farcs5 - 0\farcs8$ when using the LR5 grating and for LR6 it was $\sim 0\farcs8$ for Haro 11 and $0\farcs9 - 1\farcs3$ for ESO 338. The basic bias, flat, wavelength calibrations and sky subtractions were made with recipes from the common pipeline library (CPL) package version 5.2.0 in \texttt{esorex}. 

The bias frames were median combined with the \texttt{gimasterbias} recipe, producing master bias frames for each observing night. The \texttt{gimasterflat} recipe was then used for calibrating the fiber positioning onto the CCD. The flat frames used for each observing block were always acquired during the same night, using the Nasmyth screen (which gives better illumination in Argus mode). The \texttt{giwavecalibration} recipe subsequently performs the wavelength calibration by mapping known wavelengths from a calibration lamp. This recipe requires an existing dispersion solution as an initial guess, which is then refined. We checked the convergence of this procedure by giving the refined dispersion solution as a new initial guess and ensuring that the two refined solutions were identical for each calibration set. Finally, the \texttt{giscience} recipe applies all calibrations to the data and creates a data cube with two spatial and one spectral dimension.

Each science spectrum was reduced individually. As a sanity check, we verify the sky subtraction by measuring the continuum flux in regions far from the bright condensations, which we find to be consistent with zero. 

The subsequent reduction steps were made with the PyFITS module version 2.3.1 for Python\footnote{PyFITS is a product of the Space Telescope Science Institute, which is operated by AURA for NASA}. 
Since the science data were dithered (the telescope made small movements of roughly 0.5 to 1 arcsec between each science exposure), we created an empty x-y-$\lambda$-cube with spatial dimensions matching those of the region corresponding to the region on the sky for which we had contributions from each science exposure. The spectral data were then combined onto this grid using a weighted average intensity for each integral field unit pixel (or "spaxel"), calculated from the uncertainty values obtained from the pipeline. The dithering shifts were rounded off to whole spaxels before combining. This may thus introduce a small astrometric shift of up to 0$\farcs$25 in an individual exposure. Our final spatial resolution is therefore expected to be similar to our worst seeing of about one arcsec. 
The final data product 
has the original $0\farcs52$/lenslet spatial sampling resolution and 0.2 $\AA$ spectral sampling resolution. However, we always consider binned spaxels in a 3x3 configuration in this paper, corresponding to a bin size of $1\farcs56$.

Line profiles where fitted using the optimize.leastsq routine from the Scipy\footnote{http://www.scipy.org} version 0.7.0 Python package, which uses a Levenberg-Marquardt least-squares minimization technique.

\subsection{VLT/X-Shooter spectra}

\begin{table}
\caption{X-Shooter observing specifications}            
\label{table:1}     
\centering                     
\begin{tabular}{c c c c}    
\hline\hline               
Parameter & Haro 11B & Haro 11C & ESO 338-IG04 \\   
\hline                       
   Airmass & 1.016 & 1.012 & 1.430 \\    
   Seeing & $0\farcs7$ & $0\farcs7 - 1\farcs3$ & $0\farcs6$ \\
   Exposure time & 800 s & 680 s & 600 s \\
\hline                              
\end{tabular}
\end{table}

The X-Shooter data were obtained as part of the first science verification for the instrument between August 10 and 11 2009 (ESO ID 60.A-9433(A)). 
For the VIS arm ($\lambda \sim$ 5500 - 10200), a slit of $0\farcs9 \times 11\arcsec$ was used. For the UVB arm ($\lambda \sim$ 3000 - 5600), the slit was $1\arcsec \times 11\arcsec$. The resolving power $\mathcal{R}$
is 8800 and 5100 for the VIS and UVB arms, respectively.

The spectra were reduced with the X-Shooter pipeline v. 1.3.7 using \texttt{esorex} v. 3.9.0. Standard settings were used in the physical model mode, using the \texttt{xsh\_scired\_slit\_nod} recipe to perform reduction, sky subtraction and extraction of the science frames. 

These data are also analyzed by \citet{guseva-2012}, but we have included different reduction steps and we do not discuss the data from the NIR arm. With the large wavelength range of these data, we can examine the effect of the underlying photospheric Balmer absorption on the H$\alpha$ emission feature. It is strongest in ESO 338, but negligible for our discussion in all cases.

\section{Results}

   \begin{figure*}
   \centering
   \includegraphics[width=16cm]{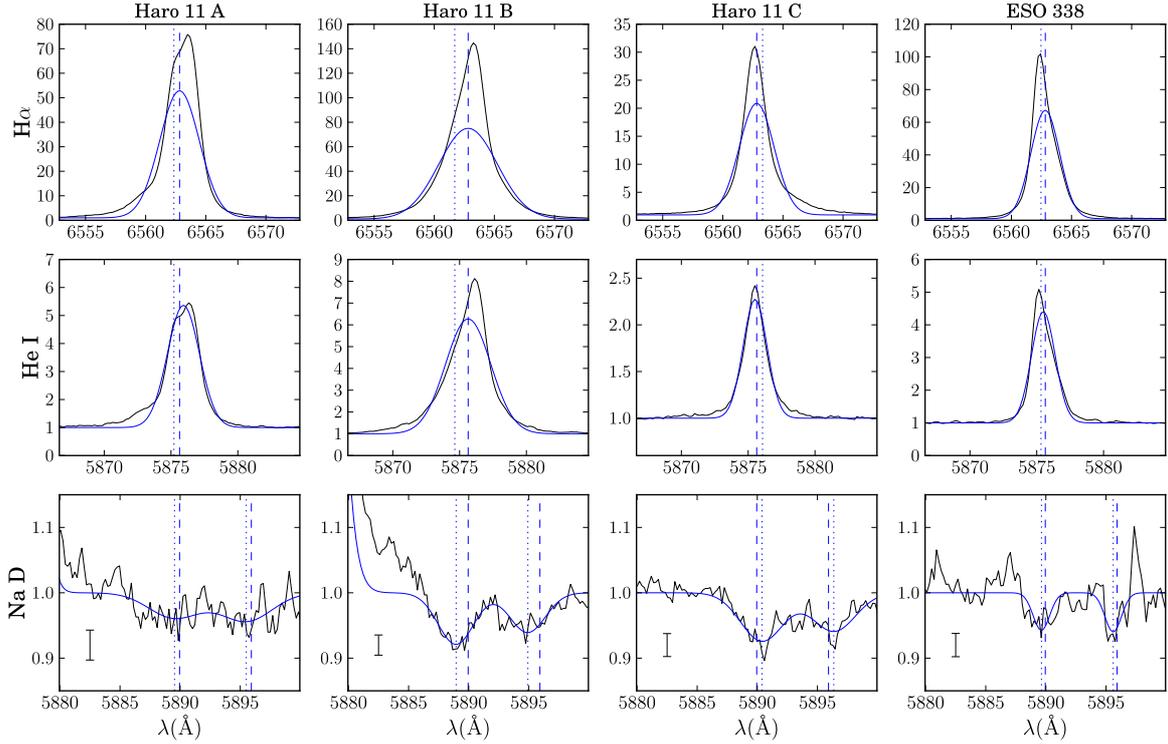}
   \caption{H$\alpha$, He {\sc i} (5875.6) and Na D line profiles from the VLT/FLAMES data.  The blue gaussians are a simple fit to the emission or absorption lines. In the case of Haro 11 knot A, the best fit is shown in this figure but is never used.
    The vertical dashed lines show the velocity given by the H$\alpha$ gaussian fit, and the vertical dotted lines show the Na D velocity. The y-axis is normalized to the continuum. Also shown inset in the bottom row are errorbars, representing the standard deviation in the continuum from 5900 to 5950 $\AA$ which shows no emission or absorption features. Since the spectra are normalized to the continuum, these errorbars reflect the strength of the continuum in each region. }
              \label{Argus_lines}
    \end{figure*}

The Na D feature is very weak in these galaxies and we were therefore forced to bin our FLAMES spectra in the spatial dimension. Even in this case, only the very strongest continuum features show Na D absorption. This reduces our discussion of the larger fields of view of FLAMES to the selected binned apertures shown in Figure~\ref{Fig_Slits}. Except for knot A in Haro 11, these are the only regions where we can see a clear Na D absorption profile. Nevertheless, these apertures still allow us to perform a sufficiently detailed analysis in some of the most interesting Ly $\alpha$ emitting and absorbing regions.

The most important spectral features from the FLAMES binned apertures are shown in Figure \ref{Argus_lines}. The spectra are not flux calibrated and are instead shown normalized to the continuum. The continuum was in all cases fitted with a low-order polynomial across the entire observed range, which yielded a good fit. The H$\alpha$ and He {\sc i} (5875.64 $\AA$) lines both show strong multi-component features, as is evident from the deviation from a simple gaussian. Note that these features are very similar in both emission lines. We verify that the asymmetry is not an instrumental feature by examining the shape of skylines and of calibration lamp frames. The measured velocities of the different spectral components are listed in Table~\ref{Table_XSHvel}.

We have also attempted multi-component fits to the H$\alpha$ emission lines (not shown). In all cases, we identify two strong components in the H$\alpha$ lines: One narrow "main" component and one broader but weaker component. In the case of Haro 11 B and ESO 338, the velocity of the broad component is consistent with the velocity of the main component, within the uncertainties. For Haro 11 A the narrow (broad) component has a velocity of $6242\pm7$ ($6189\pm10$), and it is clear that the profile is very complex, but we have no good Na D velocity to compare it to. In the case of Haro 11 C the weak, broad component is centered roughly at the same velocity as the Na D profiles. In this case, the contribution to the Ly $\alpha$ flux from this component is therefore not only expected to be weaker intrinsically but should also suffer a larger resonance effect. In each case, these secondary components do not change the velocity of the main component by more than a few kilometers per second and for simplicity we have chosen not to include them in the further discussion of this paper. Our goal is not to perfectly reproduce the line shapes but to analyze the velocities of the main constituents of the galaxies. We note that the components we measure in Haro 11 agree well with the recent results from \citet{james-2013}.

We also attempted a simple wavelength centroid fit, numerically defined as $\Sigma \lambda f_{\lambda} / \Sigma f_{\lambda}$ across the continuum subtracted emission lines. The result is perfectly consistent with the numbers we use in this paper, with a maximum shift in velocity of 5 km/s.

The X-Shooter spectra are shown for comparison in Figure~\ref{Fig_XSHLines} (see also Section \ref{sect:stellarNaD}). Note that single gaussian fits appear to reproduce the line shapes better for these data, but this is mainly due to a lower spectral resolution and signal to noise. 

We see the velocity of emission lines from ionized gas are the same within each region we measure, independent of the species used. In the same region, the velocities as measured by the emission lines of H$\alpha$ ($\lambda$ 6562.8 $\AA$), He {\sc i} ($\lambda$ 5875.6 $\AA$), [O {\sc i}] ($\lambda \lambda$ 6300.2, 6363.9 $\AA$) [N {\sc ii}] ($\lambda \lambda$ 6548.1, 6583.6 $\AA$) and [S {\sc ii}] ($\lambda \lambda$ 6716.5, 6730.7 $\AA$) are all the same, within uncertainties.   

In general, the Na D features are weak in both galaxies. For Haro 11, we measure an equivalent width for the doublet of $EW_{\rm NaD} = -0.42 \pm 0.06 \AA$ for knot C, $ -0.36 \pm 0.06 \AA$ for knot B, and $ -0.15 \pm 0.04 \AA$ for ESO 338. For Haro 11 A we estimate an upper limit of $ -0.15 \AA$. These values are estimated purely from the FLAMES spectra. 
The ratio between the EW of the separate lines of the doublet appears relatively close to unity (cf. Table \ref{Tab:NaD_ratios}), which indicates a high optical depth in the individual clouds that fall on sight lines where absorption is occurring. Note, however, that this is not the same as the average optical depth along all sight lines in the aperture, which is clearly smaller. The Na D line ratio in the optically thin case is 2:1.
We present in Table \ref{Tab:NaD_ratios} the estimated optical depths at the line center of Na D, derived from the line ratio (using the conversion in Table 2.1 from \citep{spitzer1968}), which are all found to be optically thick. In this regime, the covering fraction can be easily estimated as $C_f = 1 - I_{5890}$ where $I_{5890}$ is the residual intensity in the blue Na D line. Note however, that the residual intensity may be low in narrow, unresolved components in our spectrum and that we therefore measure an artificially stronger intensity. We will come back to this point later in the discussion. With this simple assumption however, we estimate as a lower limit the covering fraction of Na D to be $\sim 10~\%$ for Haro 11 knots B and C, and roughly 5 \% for ESO 338. 

\begin{table}
\caption{Na D properties and dust extinction}         
\label{Tab:NaD_ratios}    
\centering                    
\begin{tabular}{c c c c}   
\hline\hline          
Parameter & Haro 11B & Haro 11C & ESO 338-IG04 \\  
\hline                     
  Na D doublet EW ($\AA$) & $-0.36\pm0.06$ & $-0.42\pm0.06$ & $-0.15\pm0.04$ \\
   Na D 5890/5896 ratio & $1.15\pm0.22$ & $1.14\pm0.18$ & $0.94\pm0.47$ \\      
   Na D optical depth & $> 2$ & $> 2$ & $\gtrsim 1.6$  \\
   Na D mean shift (km/s) & $-44 \pm 13$  & $32 \pm 12$  & $-15 \pm 16$  \\
   Nebular E(B-V) & 0.42$^{a}$ & 0.48$^{a}$ & $<0.1^{b}$ \\
\hline                     
\end{tabular}
\tablefoot{References: a) \citet{hayes-2007}, b) \citet{bergvall-ostlin2002} }
\end{table}

For Haro 11, we detect no measurable Na D absorption toward knot A. This is probably due to the low continuum brightness in this region which is necessary for observing absorption lines, combined with a low covering fraction of neutral gas. For knot B, we are able to measure the Na D velocity and find a moderate blueshifted velocity (compared to the ionized gas) of 44$\pm$13 km/s. In knot C the absorption is stronger and we find a redshifted velocity of 32$\pm$12 km/s.

In ESO 338, we are only able to measure the Na D velocity in the brightest knot (knot A), were both the continuum and the Ly $\alpha$ emission is strongest.
We find no evidence for a shift in the velocity, with the best fit giving -15$\pm$16 km/s. \citet{ostlin-2007} estimated a shift of circa 20 km/s from UVES spectra with a lower signal to noise. Our estimate of the stellar versus nebular velocity agrees well with the analysis of \citet{cumming-2008}, where the stellar component (as measured by the Ca {\sc ii} triplet ($\lambda\lambda$ 8498, 8542, 8662 $\AA$) shows the same velocity as the nebular gas, with a value of 2860$\pm$4 km/s.

\section{Possible stellar contamination of Na D}\label{sect:stellarNaD}

   \begin{figure}
   \centering
   \includegraphics[width=9cm]{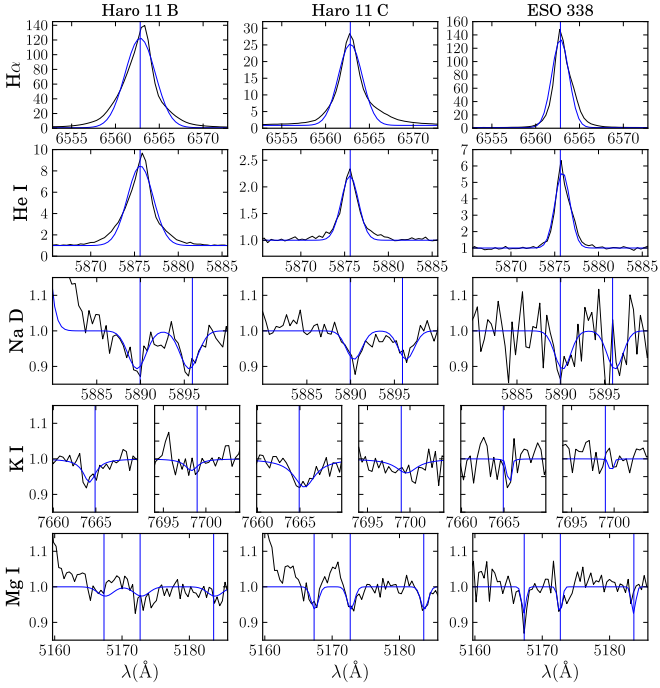}
   \caption{H$\alpha$, He {\sc i} (5875.6), Na D, K {\sc i} and Mg {\sc i} line profiles from the VLT/X-Shooter data. The blue gaussians are a simple fit to the emission or absorption lines. The vertical lines show the systemic velocity as given by the H$\alpha$ gaussian fit. The y-axis is normalized to the continuum. }
              \label{Fig_XSHLines}
    \end{figure}

A challenge with using Na D to measure neutral gas flows is that the stars in the galaxy may also exhibit the absorption lines. In this section we demonstrate that the Na D we measure is primarily of interstellar origin.

Na D is prominent in spectra of cool stars: K- and M-type giants and supergiants. There are several ways of disentangling the stellar and interstellar sodium. One of the more common and robust methods make use of the Mg {\sc i} b-band triplet ($\lambda \lambda$ 5167.32, 5172.68, 5183.60 $\AA$). It has a very similar ionization potential as Na, and the two elements are formed under similar conditions in stars. However, the triplet forms between excited states and is not found in the cold, neutral medium. Therefore, any detection of Mg {\sc i} b must come from stellar atmospheres. 

Thanks to their similar origins, the equivalent widths of the Mg b triplet and the stellar Na D doublet appear to be strongly correlated. However, any ground-based spectra of nearby stars are always contaminated by telluric Na D in emission, which makes the relative ratio difficult to measure.  \citet{rupke-2002} compile measurements of Galactic globular clusters and mostly nonactive galaxies \citep{bica-alloin1986} and nuclei of nearby galaxies \citep{heckman-1980}. They find the equivalent widths of the purely stellar features to be correlated as $EW_{\rm Na D} \sim 0.5~EW_{\rm Mg~b}$, with a possible intrinsic scatter of $\gtrsim 25 \%$.
\citet{heckman-2000} combine the data from \citet{heckman-1980} with stellar library data from \citet{jacoby-1984} and estimate $EW_{\rm Na D} \sim 0.75~EW_{\rm Mg~b}$, which seems to agree with their own K giant spectra to within 0.10 dex. 
\citet{schwartz-martin2004} also make a fit to the data by \citet{jacoby-1984} to find $EW_{\rm Na D} = 0.40 \pm 0.05~EW_{\rm Mg~b}$. \citet{martin2005} analyze Keck II/ESI spectra of A, F, G and K dwarfs and giants and conclude $EW_{\rm Na D} = 1/3~EW_{\rm Mg~b}$ and that the stellar contribution was consistently less than 10 \% in their study of 18 ULIRGs.  
\citet{sato-2009} investigate Na D absorption in 493 spectra from the AEGIS survey, and find $EW_{\rm Na D} = 0.40~EW_{\rm Mg~b}$ well describes the purely stellar boundary in their sample (see their Fig.~1).
Finally, \citet{chen-2010} used Sloan Digital Sky Survey (SDSS) spectra from young disc galaxies and estimate that an average $\sim 80\%$ of the Na D absorption arises in stellar atmospheres. Their estimate of the stellar contribution is based on fitting and subtracting a stellar population synthesis model to the continuum and absorption features, and they emphasize that their spectral model sometimes shows stronger Na D absorption than the actual data. They conclude that the most likely explanation is that Na D is sometimes seen in emission in their data, predominantly from face-on galaxies with low dust attenuation, and that a stronger stellar contamination is expected in the spectra of normal star-forming galaxies as opposed to that in young stars comparable to the dominant stellar population in our two galaxies here.

We note that Na D is a resonance absorption feature, and therefore absorbed photons will be re-emitted in a random direction. This can create a diffuse, low surface brightness emission component. \citet{prochaska-2011} model the effect of re-emitted light and conclude that in the extreme case that all of this light eventually escapes and is caught within the aperture, it can significantly reduce the observed equivalent width. We expect this effect to be very small in our analysis, however, since we are looking only at narrow slits and small effective apertures (1.5 arcsec corresponds to roughly 0.6 kpc in Haro 11 and even less in ESO 338). 

An alternative to measuring outflows with Na D is the K {\sc i} doublet ($\lambda\lambda$ 7664.91, 7698.97 $\AA$). With similar properties and a similar ionization potential of 4.34 eV (compared to 5.14 eV for Na D), these lines are expected to behave much in the same way \citep[see e.g.][]{kemp-2002}. They are however very seldom used, because they are weaker than Na D due to a lower relative abundance but also because they often overlap with atmospheric ${\rm O_2}$ absorption bands. 

Here, we use our X-shooter data to estimate the stellar contamination. Due to the comparatively low signal-to-noise and lower resolution of the X-Shooter data, the equivalent widths are not as easily determined as for the Argus data.  However, we can still use our spectra in Figure \ref{Fig_XSHLines} to see that it is only in Haro 11 C that we can clearly distinguish the Mg b feature, although it is very weak. We thus let the upper limit from Haro 11 C represent the upper limit on the stellar contamination for all our results. If the EW of Mg b would be larger than this limit in the other regions this would be readily seen in Figure~\ref{Fig_XSHLines}. We estimate the EW of this feature to be roughly -0.2 $\AA$, i.e. about half the equivalent width of Na D. 

Based on the discussion above we thus estimate the stellar contribution to be on the order of 25-30\%, but we note that it may be as high as 50\% in the extreme case. A significant contribution to the Na D profile at the stellar velocity (which is consistent with the H-alpha velocity in our galaxies) would decrease the average offset velocities that we measure. The absolute value of our Na D velocities may therefore be somewhat lower than the purely nebular Na D velocity.

\begin{table}
\caption{Best fit observed velocities, in km/s. } 
\label{Table_XSHvel}   
\centering                 
\begin{tabular}{c c c c c} 
\hline\hline             
Line & Haro11A & Haro 11B & Haro 11C & ESO 338-IG04 \\   
\hline                   
   H$\alpha$  & 6227$\pm$10 & 6146$\pm$5 & 6126$\pm$7 & 2859$\pm$6 \\    
   He {\sc i} & 6242$\pm$9 & 6147$\pm$6 & 6121$\pm$9 & 2856$\pm$11 \\
   Mg {\sc i} & ...        & 6162$\pm$20 & 6128$\pm$15 & 2853$\pm$20 \\
   Na D       & ...        & 6102$\pm$9 & 6158$\pm$8 & 2844$\pm$14 \\
   K {\sc i}  & ...        & 6123$\pm$20 & 6146$\pm$20 & 2830$\pm$20 \\
\hline                                  
\end{tabular}
\tablefoot{ Mg {\sc i} and K {\sc i} are measured only from VLT/X-Shooter and the uncertainty is therefore larger. The other lines were measured in both VLT/FLAMES and VLT/X-Shooter and this represents the weighted average.}
\end{table}

\section{Discussion}

We have measured outflow/inflow velocities and covering fractions of the neutral ISM in front of three bright star-forming condensations, two of which show Ly $\alpha$ emission (ESO 338 A, and Haro 11 C) and one shows Ly $\alpha$ in absorption (Haro 11 B). Further to the information measured in this programme, we also compile measurements of the nebular dust attenuation in these regions from our previous investigations: \citet{hayes-2007} for Haro 11, \citet{bergvall-ostlin2002} for ESO 338, see Table~\ref{Tab:NaD_ratios}. 

\subsection{Comments on the individual regions}

\paragraph{Emission -- ESO 338 knot A}

For ESO338-IG04, we find that the velocity of the neutral gas (as estimated from the Na D absorption) is only very slightly blueshifted (or even static) compared to the ionized gas. This would rule out the outflow scenario as an explanation for the observed direct Ly $\alpha$ escape from this region. However, we note that H$\alpha$ from the central region (knot A) of ESO 338 is dominated by a large H {\sc ii} shell \citep[e.g.][]{bergvall-ostlin2002}. The UVES spectra analyzed by \citet{ostlin-2007} show a multi-component feature in the [O {\sc iii}] ($\lambda 5007 \AA$) and H$\alpha$ lines towards knot A, which they interpret as an expanding bubble at $\sim$~40 km/s. Presumably this shell does consists of outflowing ionized gas from stellar winds and supernovae, yet we measure a low outflow velocity in H {\sc i}. If there is indeed an expanding ionized bubble in ESO 338, it is possible that the H$\alpha$ feature is dominated by emission on the side of the bubble facing our way. This would cause us to measure a more blueshifted velocity for the ionized gas and reduce our inferred neutral gas velocity. However, the bubble seems to be optically thin, which would mean that the effect is very small, and also the stellar component in knot A appears to have the same velocity as the ionized gas \citep{cumming-2008}.

\paragraph{Emission -- Haro 11 knot C}

For Haro 11, we find a small redshifted velocity of Na D in front of knot C, indicating infall of cold gas. HST imaging shows Ly $\alpha$ emission from this region, which could partially be explained by this velocity difference. However, it is in the opposite direction to an outflow, and the velocity difference is not particularly strong. 

There is evidence for a Ly $\alpha$ / H$\alpha$ ratio higher than the recombination case B value in both ESO 338 A \citep{ostlin-2009} and Haro 11 C \citep[if extinction is taken into account;][]{atek-2008}. This could be due to Ly $\alpha$ photons actually suffering less attenuation while scattering on the surface of cool dusty clumps than the continuum photons \citep{neufeld1991}, although this scenario seems unlikely based on recent simulations \citep[][Duval et al. 2012, submitted]{laursen-2012}. It is also possible that an extinction correction based on a clumpy dust distribution model rather than a uniform dust screen would give a result consistent with recombination values \citep[see e.g.][]{scarlata-2009}.

\paragraph{Absorption -- Haro 11 knot B}

For knot B, we find a moderate blueshifted velocity. Combined with a low covering fraction $\sim 10~\%$, Ly $\alpha$ should escape more easily from knot B. However, our HST UV imaging shows that Ly $\alpha$ is strongly absorbed in this region. 

There is some debate as to whether the dust extinction is higher in knot B than in knot C. Judging from HST images, there do seem to be more dust clouds near and around knot B and based on the X-Shooter data, knot C seems to have a lower extinction \citep{guseva-2012}. However, the H$\alpha$ and H$\beta$ images from which we derive the E(B-V) values presented in Table~\ref{Tab:NaD_ratios} give approximately the same level of extinction in both knots \citep{hayes-2007,atek-2008}. If the medium is highly clumpy it is likely that the extinction is varying, even within these small regions. Different results are thus likely due to different aperture sizes and positions, as well as different techniques for measuring the extinction probing varying optical depths.

\paragraph{Absorption -- Haro 11 knot A}

The signal-to-noise and the equivalent width of Na D are too small for us to safely attempt a measurement toward knot A.
This is partly because the optical continuum emission near Na D is weaker than in the other knots, but it still indicates a low covering fraction of neutral gas. Knot A exhibits both H $\alpha$ and UV continuum emission from ionized regions, but no significant Ly $\alpha$ radiation appears to escape from them.

\subsection{Interpretation of the covering fractions, optical depths and velocities}

In the three regions that we have identified for study in this article, we have compiled measurements of H {\sc i} covering fractions, kinematics, and dust contents. 

In both knot A and B in Haro 11, absorption of Ly $\alpha$ is seen despite a low covering fraction. However, the covering fractions that we estimate from the Na D profiles likely only serves as lower limits, since there may be very narrow, unresolved components with lower residual intensity in our spectra. We note that the O {\sc i} and Si {\sc ii} lines in the GHRS spectra of \citet{kunth-1998} are considerably stronger, indicating a high covering fraction of neutral gas, consistent with the strong Ly $\alpha$ absorption on the blue side of the line. The resolution of GHRS ($\mathcal{R} \sim 16000$) is not much higher than that of our Argus Na D spectra ($\mathcal{R} = 11800$), and would likely not explain this difference. We note also that other Na D absorption spectra in the literature often show considerably stronger lines \citep[e.g.][]{heckman-2000,martin2005}. Our conclusion is that the Na D covering fraction is indeed low in these regions. The difference from the previous measurement might in part be explained by the different apertures pointing at different sightlines in an inhomogenous ISM. The difference may also come from the lower photoionization threshold of Na D of 5.14 eV. It is thus possible that the column density of hydrogen is high enough overall to absorb Ly~$\alpha$ but low enough that we detect Na D only in the densest regions. The interstellar medium is known to often be very patchy, and it would not be surprising in these two galaxies with their turbulent pasts. Since Na D has to be shielded from ionizing radiation by dust \citep{chen-2010,murray-2007}, it likely only exists in the densest, coolest clumps. 

For the optical depths that we find from the Na D line ratio, we can put a lower limit on the column density of the gas where the absorption arises. We use the relation 

\begin{equation}
{\rm N(Na~\textsc{i})} = \frac{\tau_{1}  b }{1.497 \times 10^{-15}  \lambda_{1}  f_{1}}
\end{equation}

from \citet{spitzer1978}, where $\tau_1$ is the central optical depth, $b$ is the Doppler parameter (in km/s), $\lambda_1$ is the vacuum wavelength in $\AA$ and $f_1$ is the oscillator strength = 0.3180, all for the weaker (red) line. With a Doppler parameter $b = FWHM / (2\sqrt{\ln{2}})$ of $\gtrsim$~60 km/s for the Na D in ESO 338, a lower limit on the Na {\sc i} column density is roughly ${\rm N(Na~{\textsc i})} \gtrsim 4\times10^{13} {\rm cm}^{-2}$. Converting this to a hydrogen column density of course depends heavily on the Na / H abundance ratio, but using very conservative estimates based on the conversions given in \citet{rupke-2002, murray-2007} this corresponds to at least N(H {\sc i}) $> 10^{20} {\rm cm}^{-2}$. Ly~$\alpha$ becomes optically thick and resonantly scatters on neutral hydrogen already at H {\sc i} column densities around $10^{13-14} {\rm cm}^{-2}$. It would take an enormous departure from these assumptions to make the clouds optically thin. In these dense clouds, the Ly~$\alpha$ photons would see upwards of a million optical depths, and the clouds would be self-shielding. Still, the covering fraction of these dense clouds where we see Na D is only $\sim 0.1$. Our interpretation is that the ISM in these galaxies is inhomogeneous and likely consists of column densities between these extremes. Ly~$\alpha$ photons would then escape through regions or patches with little neutral gas, and be blocked by gas at slightly higher column densities, while we measure our velocities and covering fractions in the very densest clumps. 

For Haro 11, a picket-fence model of the ISM agrees well with the detection of the Lyman continuum (Ly C, $< 912 \AA$) escape \citep{bergvall-2006,leitet-2011}, for which {\em only} direct paths are viable \citep[see e.g.][]{heckman-2001,zastrow-2011}.
An indirect method of estimating Ly C escape through the residual in the C {\sc ii} ($\lambda$ 1036) line points to that ESO 338 is also a strong candidate for Ly C escape (Leitet et al. 2012, submitted), but unfortunately the redshift is too low for a direct measurement. 

The question still remains why, between knot B and C in Haro 11 with their relatively similar nebular dust content, Ly $\alpha$ is absorbed in knot B, which shows the {\em larger} kinematic offset. We hypothesize on the existence of diffuse remnant H {\sc i} gas with a high covering fraction towards knot B. 

Our measurements of the neutral gas velocities in Haro 11 B agree rather well with the results in \citet{kunth-1998} where an outflow of $\sim$60 km/s measured by low ionization state absorption lines in the UV was seen. However, the same spectra show Ly $\alpha$ in emission which would imply that some of the emission from knot C is included in the aperture as well. 
It is unfortunately not clear exactly where in the galaxy the HST/GHRS aperture was placed \citep[see discussion in][]{hayes-2007}. Most likely, it was in between the knots, and it is quite likely that the different components are mixed in the spectra. Indeed, \citet{kunth-1998} see very broad absorption features (spanning roughly 200 km/s), indicating multiple absorption components. 

We note that a previous measurement of the neutral gas velocity in ESO 338 was made with HST/STIS in \citet{schwartz-2006}, which agrees well with our observations. The LIS absorption features are weak and appear close to the systemic velocity. Unfortunately, the available HST/STIS spectrum for ESO 338 is not optimized for exploring the Ly $\alpha$ emitting region that we investigate in this paper. In light of our results, it would be very interesting to explore both Haro 11 and ESO 338 with multiple HST/COS and HST/STIS pointings to investigate the Ly $\alpha$ emission and absorption line profile in these galaxies in unprecedented spatial and spectral detail, which can be achieved by combining the two instruments.

\section{Conclusions}
We have presented the first spatially resolved measurements of the sodium doublet (Na D) in the two nearby Ly $\alpha$ emitting galaxies Haro 11 and ESO 338-IG04. Our results can be summarized as follows:

\begin{itemize}

\item 
We find an outflow of neutral gas from knot B (which shows strong Ly $\alpha$ absorption) and slow infall towards knot C in Haro 11. In ESO 338-IG04 we find a slow or static interstellar medium. In the two latter cases, Ly $\alpha$ is seen in emission. 

The velocities that we find for the neutral gas are not what we would have expected from standard Ly $\alpha$ escape scenarios. Typically, a strong outflow is assumed to allow Ly $\alpha$ to escape more easily. 

\item
From the Na D line profiles we measure relatively high optical depths but small covering fractions of Na D ($\sim 10 \%$). We estimate the minimum column density of Na D corresponding to our limits on the optical depth and find that Na D is likely only detected in the very densest clumps, where ${\rm N(H~{\textsc i})} > 10^{20} {\rm cm^{-2}}$. Since Ly $\alpha$ is affected by resonant scattering already at column densities of ${\rm N(H~{\textsc i})} \sim 10^{13-14} {\rm cm^{-2}}$, it is likely that the direct Ly $\alpha$ escape seen in e.g. knot C in Haro 11 and from ESO 338 is due to a picket-fence scenario 
where the interstellar medium is highly inhomogeneous and consists of both dense, neutral clumps as well as ionized gas along our lines of sight.

\item 
We see a larger kinematical offset in Haro 11 knot B than C, yet B shows strong Ly $\alpha$ absorption in contrast to the emission from C. Given the relatively similar nebular dust content of the two knots, we hypothesize on the existence of a diffuse remnant H {\sc i} component with a high covering fraction towards B, and a possible perpendicular outflow from C.

\end{itemize}

\begin{acknowledgements}

We would like to thank Matthew Lehnert for a fruitful discussion regarding the stellar contamination problem. M.H. received support from Agence Nationale de la recherche bearing the reference ANR-09-BLAN-0234-01. G{\"O} is a Royal Swedish Academy of Sciences Research Fellow, supported by a grant from the Knut and Alice Wallenberg foundation, and also acknowledges support from the Swedish Research Council and the Swedish National Space Board.  J.M.M.H. is funded by Spanish MINECO grants AYA2010-21887-C04-02 (ESTALLIDOS) and AYA2012-39362-C02-01. 

\end{acknowledgements}

\bibliographystyle{aa} 
\bibliography{sandbergetal2013.bib} 
\end{document}